\documentclass[useAMS]{mn2e}

\usepackage{latexsym,graphicx}

\usepackage{color}
\usepackage{bm}

\newcommand\kms{{\rm\,km\,s^{-1}}}
\newcommand\msun{\rm\,M_\odot}

\newcommand\rsun{\rm\,R_\odot}
\newcommand\myr{\msun \, {\rm yr}^{-1}}
\newcommand{\MC}{\multicolumn}
\newcommand\hii{H\,{\sc ii} \,}

\title[Modelling interstellar structures around Vela\,X-1]{Modelling interstellar structures around Vela\,X-1}
\author[V.V.Gvaramadze et al.]
        {V. V.~Gvaramadze,$^{1,2,3}$\thanks{E-mail: vgvaram@mx.iki.rssi.ru} D. B.~Alexashov,$^{2,4}$ O. A.~Katushkina$^{2}$ 
        \newauthor and A. Y.~Kniazev$^{1,5,6}$ \\
        $^{1}$Sternberg Astronomical Institute, Lomonosov Moscow State University, Universitetskij Pr. 13, Moscow 119992, Russia\\
        $^{2}$Space Research Institute, Russian Academy of Sciences, Profsoyuznaya 84/32, 117997 Moscow, Russia \\
        $^{3}$Isaac Newton Institute of Chile, Moscow Branch, Universitetskij Pr. 13, Moscow 119992, Russia \\
        $^{4}$Institute for Problems in Mechanics, prosp. Vernadskogo 101, block 1, Moscow, 119526, Russia \\
        $^{5}$South African Astronomical Observatory, PO Box 9, 7935 Observatory, Cape Town, South Africa \\
        $^{6}$Southern African Large Telescope Foundation, PO Box 9, 7935 Observatory, Cape Town, South Africa \\
        }
\begin{document}

\date{Accepted 2017 November 24. Received 2017 November 24; in original form 2017 October 26}

\maketitle

\label{firstpage}

\begin{abstract}
We report the discovery of filamentary structures stretched behind the bow-shock-producing high-mass X-ray binary Vela\,X-1 using
the SuperCOSMOS H-alpha Survey and present the results of optical spectroscopy of the bow shock carried out with the Southern African
Large Telescope (SALT). The geometry of the detected structures suggests that Vela\,X-1 has encountered a wedge-like layer of enhanced
density on its way and that the shocked material of the layer partially outlines a wake downstream of Vela\,X-1. To substantiate 
this suggestion, we carried out 3D magnetohydrodynamic simulations of interaction between Vela\,X-1 and the layer for three limiting 
cases. Namely, we run simulations in which (i) the stellar wind and the interstellar medium (ISM) were treated as pure 
hydrodynamic flows, (ii) a homogeneous magnetic field was added to the ISM, while the stellar wind was assumed to be
unmagnetized, and (iii) the stellar wind was assumed to possess a helical magnetic field, while there was no magnetic field 
in the ISM. We found that although the first two simulations can provide a rough agreement with the observations, only the third
one allowed us to reproduce not only the wake behind Vela\,X-1, but also the general geometry of the bow shock ahead of it. 
\end{abstract}


\section{Introduction}
\label{sec:int}

A wind-blowing star moving supersonically through the interstellar medium (ISM) generates a bow shock in the upwind direction. 
The shape of the bow shock depends on the space velocity of the star (the higher the velocity, the smaller the opening angle of 
the bow shock), while the strength of the shock (i.e. the degree of compression of the shocked ISM) reduces from apex to flanks. 
Correspondingly, the bow shock appears as an arcuate structure with the maximum emission intensity at its nose part. Numerous
examples of such structures were detected with infrared space telescopes (e.g., van Buren, Noriega-Crespo \& Dgani 1995; 
Gvaramadze \& Bomans 2008; Gvaramadze et al. 2011b; Peri et al. 2012; Peri, Benaglia \& Isequilla 2015; Kobulnicky 
et al. 2016), some of which have counterparts in optical and other wavelengths. 

The discovery of stellar wind bow shocks has triggered many theoretical studies of this phenomenon. 
Baranov, Krasnobaev \& Kulikovskii (1970) and Wilkin (1996) described (respectively, numerically and analytically) the shape of 
isothermal bow shocks by using the thin-layer approximation. The obtained solution provides a good description of bow shocks
produced by cool (e.g. asymptotic giant branch or red supergiant) stars, moving through the space with moderate velocities. These 
bow shocks are subject of various types of instabilities (e.g. Dgani, Van Buren \& Noriega-Crespo 1996; Wareing, Zijlstra \& 
O'Brien 2007a; Mohamed, Mackey \& Langer 2012), which are manifested in the ragged appearance of the bow shocks themselves as well 
as of tails behind them (e.g. Wareing, Zijlstra \& O'Brien 2007b; Sahai \& Chronopoulos 2010). 

In bow shocks produced by hot and/or fast-moving stars the thickness of the layers of shocked ISM and stellar wind could be 
comparable with the characteristic scale of the bow shock itself (e.g. van Buren 1993; Raga et al. 1997; Comer\'on \& Kaper 1998), 
so that the only way to describe their structure is through the numerical modelling. This approach to study bow shocks allows 
one to take into account various factors which could affect or even determine their structure (e.g. van Marle et al. 2011; 
Mackey et al. 2012; Decin et al. 2012; Meyer et al. 2014a,b; Acreman, Stevens \& Harries 2016). One of these factors is the 
interstellar magnetic field, whose inclusion in the models makes it possible to better understand the general and/or fine structure 
of bow shocks (van Marle, Decin \& Meliani 2014; Meyer et al. 2017; Katushkina et al. 2017, 2018). In this paper, we show that 
the shape of some bow shocks could also be affected by the stellar magnetic field.

Supersonic motion is typical of the so-called runaway stars, which got their high space velocities either because of dynamical 
few body encounters in parent star clusters (Poveda, Ruiz \& Allen 1967) or disruption of binary systems following supernova 
explosion of one of the binary components (Blaauw 1961; Boersma 1961). These high-velocity stars move thought the Galactic disk 
and interact with density inhomogeneities on their way. Signatures of this interaction might remain for a long time behind the 
runaway stars in the form of elongated bubbles or wakes of shocked gas, whose geometry is defined by the density distribution in 
the local ISM, the stellar mass-loss history and instabilities at the wind/ISM interface. The best examples of such structures are 
those associated with the progenitor stars of SN\,1987A (Wang, Dyson \& Kahn 1993) and supernova remnant RCW\,86 (Gvaramadze et al. 
2017), the pulsar PSR\,B2223+65 (Cordes, Romani \& Lundgren 1993) and Mira (Martin et al. 2007).

Vela\,X-1 is a runaway high-mass X-ray binary (HMXB; van Rensbergen, Vanbeveren \& De Loore 1996; Kaper et al. 1997) and one of 
two HMXBs with detected bow shocks (the second one is 4U\,1907+09; Gvaramadze et al. 2011a). The bow shock produced by Vela\,X-1 
was discovered by Kaper et al. (1997) using narrowband H$\alpha$ imaging. The symmetry axis of the bow shock and the direction of 
motion of Vela\,X-1 (derived from proper motion measurements) both indicate that this HMXB is running away from the Galactic plane, 
probably from the Vela\,OB1 association (van Rensbergen et al. 1996; Kaper et al. 1997; see also Section\,\ref{sec:dis}).

In this paper, we report the discovery of large-scale filamentary structures behind Vela\,X-1 using the SuperCOSMOS H-alpha Survey 
(SHS; Parker et al. 2005). We interpret them as the result of interaction of the stellar wind with the inhomogeneous ISM during 
which Vela\,X-1 has entered in a region of enhanced density. In Section\,\ref{sec:vela}, we review the relevant data on Vela\,X-1 
and its bow shock. In Section\,\ref{sec:neb}, we discuss the surroundings of Vela\,X-1 and present the SHS image of the filamentary 
structures behind it. In Section\,\ref{sec:obs}, we describe our spectroscopic observations of the bow shock. In Section\,\ref{sec:mod}, 
we describe our numerical 3D magnetohydrodynamic (MHD) model of the interaction between Vela\,X-1 and the local ISM. The results of 
the 3D simulations are presented and discussed in Section\,\ref{sec:dis}. 

\begin{table*}
\caption{Summary of astrometric and kinematic data on
Vela\,X-1 (see text for details).}
\label{tab:prop}
\begin{tabular}{ccccccccc}
\hline 
$d$ & $\mu _\alpha \cos \delta$
& $\mu _\delta$ & $v_{\rm r,hel}$ & $v_{\rm l}$ & $v_{\rm b}$ & $v_{\rm r}$ & $v_{\rm tr}$ &$v_*$ \\
(kpc) & (mas ${\rm yr}^{-1}$) & (mas ${\rm yr}^{-1}$) & ($\kms$) & ($\kms$) & ($\kms$) & ($\kms$) & ($\kms$) & ($\kms$) \\
\hline 
2.0 & $-$4.879$\pm$0.029 & 9.278$\pm$0.033 & $-$3.20$\pm$0.90 & $-$32.23$\pm$0.30 & 32.29$\pm$0.29 & $-$29.36$\pm$0.90 & 
45.62$\pm$0.30 & 54.25$\pm$0.55 \\ 
\hline
\end{tabular}
\end{table*}

\section{Vela\,X-1: general data}
\label{sec:vela}

The HMXB Vela\,X-1 (also V*\,GP\,Vel, HD\,77581 and HIP\,44368) is a short-period (8.96\,d; Forman et al. 1973) and 
low-eccentricity ($e\approx0.09$; Bildsten et al. 1997) eclipsing binary system composed of a blue supergiant star (B0.5\,Ia; 
Hiltner, Werner \& Osmer 1972) of mass of $\approx24 \, \msun$ and a neutron star of mass of $\approx1.9 \, \msun$ (the most 
massive neutron star known to date; van Kerkwijk et al. 1995; Quaintrell et al. 2003; see, however, Koenigsberger, Moreno \& 
Harrington 2012). Detection of cyclotron absorption features in the X-ray spectrum of Vela\,X-1 yields a magnetic field of 
$B_{\rm NS}=2.6\times10^{12}$\,G at the surface of the neutron star (Kretschmar et al. 1996). The neutron star accretes a 
fraction of the stellar wind and produces a pulsed X-ray emission with a period equal to its spin period of 283\,s (McClintock 
et al. 1976). The eclipsing nature of Vela\,X-1 implies that the orbital plane of this HMXB is almost parallel to our line of
sight. Indeed, from the duration of X-ray eclipses it was derived that the angle between the orbital plane and the plane of
the sky is $\ga70\degr$ (van Kerkwijk et al. 1995). The orbital separation of the binary of $53 \, \rsun$ is only a factor of
$\approx1.8$ larger than the radius of the B0.5\,Ia star of $\approx30 \, \rsun$ (van Kerkwijk et al. 1995), which implies that 
the neutron star is deeply embedded in the wind of its companion star and that there is a strong interaction between the wind 
of the blue supergiant and the magnetosphere of the neutron star. 

Though the binary parameters of Vela\,X-1 are known quite well, there is an ambiguity in determining the mass-loss rate, 
$\dot{M}$, and the terminal wind velocity, $v_\infty$, of the supergiant star (cf. Prinja et al. 1990; Kaper, Hammerschlag-Hensberge 
\& Van Loon 1993; Watanabe et al. 2006; Krti\v{c}ka, Kub\'at \& Krti\v{c}kov\'a 2015). In what follows, we adopt for $\dot{M}$ and 
$v_\infty$ the values derived in the recent study of Vela\,X-1 by Gim\'enez-Garcia et al. (2016): $\log(\dot{M}/\myr)=-6.2\pm0.2$ 
and $v_\infty=700^{+200} _{-100} \, \kms$.

Detection of the bow shock around Vela\,X-1 implies that this HMXB is moving supersonically with respect to the local ISM and that 
the medium is dense enough to make the bow shock observable (Huthoff \& Kaper 2002). In the absence of regular flows in the ISM 
(e.g. triggered by supernova explosions or outflows from nearby massive star clusters), the observed stand-off distance of the bow 
shock (i.e. the minimum distance from the star at which the wind ram pressure is equal to the ISM ram pressure) could be used to 
constrain the number density of the local medium, provided that the distance to Vela\,X-1 and its space velocity are known with a 
reasonable precision. 

Numerous distance estimates for Vela\,X-1 based on the photometry of the optical companion star are in a good agreement with each 
other: $d=1.9\pm0.2$ kpc (Sadakane et al. 1985), 2.2$\pm$0.2 kpc (Coleiro \& Chaty 2013), 2.0$\pm$0.2 kpc (Gim\'enez-Garcia et al. 
2016), as well as with the parallactic distance from the Gaia first data release (DR1; Gaia Collaboration et al. 2016a,b) of 
$1.85^{+1.85} _{-0.62}$ kpc. In what follows, we adopt a distance of 2 kpc to Vela\,X-1. At this distance, Vela\,X-1 is separated 
from the Galactic plane by $\approx140$ pc.

To derive the space velocity of Vela\,X-1, we use the proper motion measurements, $\mu _\alpha \cos \delta$ and $\mu _\delta$, for 
this system from the {\it Gaia} DR1 catalogue (Gaia Collaboration et al. 2016a,b). The heliocentric radial velocity of Vela\,X-1, 
$v_{\rm r,hel}$, is taken from the Ninth Catalogue of Spectroscopic Binary Orbits (Pourbaix et al. 2004). Using the Solar 
galactocentric distance $R_0 = 8.0$ kpc and the circular Galactic rotation velocity $\Theta _0 =240 \, \kms$ (Reid et al. 2009), 
and the solar peculiar motion $(U_{\odot},V_{\odot},W_{\odot})=(11.1,12.2,7.3) \, \kms$ (Sch\"onrich, Binney \& Dehnen 2010), we 
calculated the peculiar transverse velocity $v_{\rm tr}=(v_{\rm l}^2 +v_{\rm b}^2)^{1/2}$, where $v_{\rm l}$ and $v_{\rm b}$ are, 
respectively, the velocity components along the Galactic longitude and latitude, the peculiar radial velocity $v_{\rm r}$, and the 
total space velocity $v_*$ of the star. For the error calculation, only the errors of the proper motion and the radial velocity 
measurements were considered. The input data and resulting velocities are given in Table\,\ref{tab:prop}.

From Table\,\ref{tab:prop} it follows that Vela\,X-1 is moving in the north direction (which is consistent with the direction 
suggested by the orientation of the bow shock; see the next section) with a transverse velocity of $\approx45 \, \kms$. Moreover, 
Vela\,X-1 is approaching us with a velocity of $\approx30 \, \kms$, and the vector of its space velocity is inclined to our line of 
sight by an angle of $\approx60\degr$.

\section{Surroundings of Vela\,X-1}
\label{sec:neb}

\begin{figure*}
\includegraphics[width=14cm,clip=0]{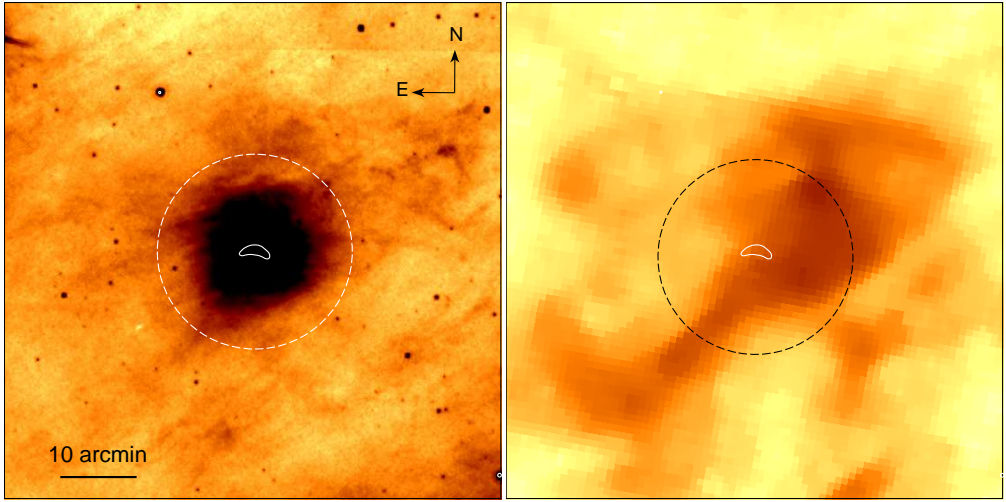}
\centering \caption{{\it WISE} 22\,$\mu$m (left-hand panel) and SHASSA continuum-subtracted (right-hand panel) images of the field 
containing Vela\,X-1. The position of the bow shock generated by Vela\,X-1 is indicated by a white contour. The area surrounded by a 
dashed circle of radius of $0\fdg2$ (centred on Vela\,X-1) corresponds to the putative Str\"omgren zone produced by Vela\,X-1. See 
text for details.} 
\label{fig:22+shassa}
\end{figure*}

Kaper et al. (1997) noted the presence of an extended (of angular radius of $\sim0\fdg2$) {\it IRAS} source at the position of 
Vela\,X-1, and suggested that it might be a Str\"omgren zone\footnote{Note that the radius of the \hii region around a moving source 
of ionizing emission is equal to the Str\"omgren radius (e.g. Gvaramadze, Langer \& Mackey 2012).} created by the ultraviolet emission 
of Vela\,X-1; see the left-hand panel in Fig.\,\ref{fig:22+shassa} for the {\it WISE} ({\it Wide-field Infrared Survey Explorer}) 
22\,$\mu$m image of the region containing Vela\,X-1 with the position of the putative Str\"omgren zone indicated 
by a dashes circle. If this infrared source is indeed an {\it ionization-bounded} \hii region, then one can derive the number density 
of the local ISM, $n_{\rm ISM}$, using the following relationship (e.g. Lequeux 2005):
\begin{equation}
n_{\rm ISM}\approx6 \, {\rm cm}^{-3} \left({R_{\rm St}\over 7 \, {\rm pc}}\right)^{-3/2}
\left({S(0)\over 4\times10^{47} \, {\rm s}^{-1}}\right)^{1/2} \, ,
\label{eqn:hii}
\end{equation}
where $R_{\rm St}$ is the Str\"omgren radius (at a distance of 2 kpc, $0\fdg2$ corresponds to $\approx7$ pc) and $S(0)$ is the ionizing 
photon luminosity, which for a B0.5\,I star is equal to $4\times10^{47} \, {\rm s}^{-1}$ (Panagia 1973). This density, as it was pointed 
out by Kaper et al. (1997), is an order of magnitude higher than what might expect at a distance of $\approx140$ pc from the Galactic 
plane, which suggests that Vela\,X-1 met a dense cloudlet on its way. One can therefore imagine to see around Vela\,X-1 a region of 
H$\alpha$ emission of the same angular size as the {\it IRAS} source (cf. Mackey, Langer \& Gvaramadze 2013).

We searched for an H$\alpha$ counterpart to the {\it IRAS} source using the Southern H-alpha Sky Survey Atlas 
(SHASSA; Gaustad et al. 2001). Inspection of this atlas revealed incoherent  structures at and around the position 
of Vela\,X-1 within an area much wider than the {\it IRAS} source. The right-hand panel of Fig.\,\ref{fig:22+shassa} 
shows the SHASSA continuum-subtracted image\footnote{The angular resolution of the SHASSA images is $\approx$0.8 
arcmin.} of the field containing Vela\,X-1 and its bow shock (indicated by a white contour). One can see that 
Vela\,X-1 is located near the edge of an elongated region of bright H$\alpha$ emission, which is stretched in the 
SE--NW direction well beyond the circle of the radius of $0\fdg2$. Later on, after the data from the SuperCOSMOS 
H-alpha Survey (SHS; Parker et al. 2005) become publicly available and thanks to the excellent angular resolution 
($\sim1$ arcsec) of these data, we found indications that the region of H$\alpha$ emission visible in the SHASSA 
image might be associated with Vela\,X-1, while the elongated structure in the SE might actually be shaped by the 
wind from this HMXB.

\begin{figure*}
\includegraphics[width=18cm,clip=0]{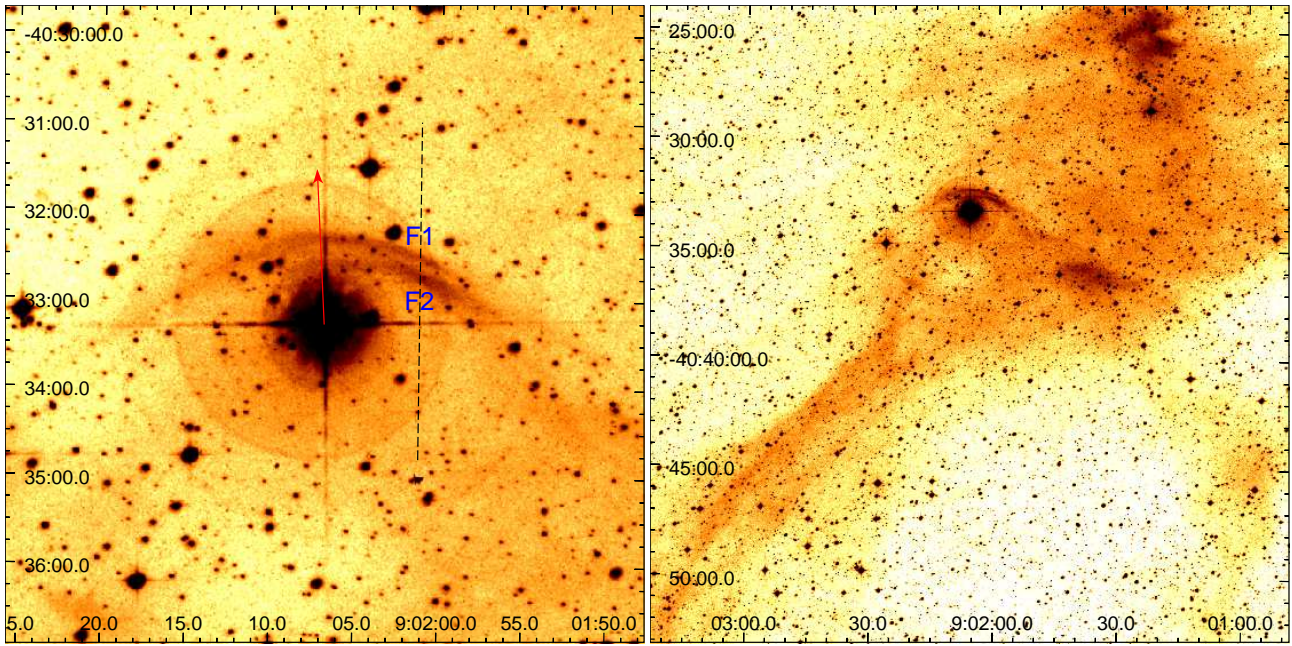}
\centering \caption{SHS H$\alpha$+[N\,{\sc ii}] images of the bow shock generated by the HMXB Vela\,X-1(left panel) and 
a $30\arcmin\times30\arcmin$ field containing Vela\,X-1 and filamentary structures behind it (right panel). The arrow and 
dashed line in the left panel show, respectively, the direction of motion of Vela\,X-1 (as follows from the Gaia proper 
motion measurement) and the orientation of the spectrograph slit. The slit crosses two filaments in the bow shock, labelled 
F1 and F2 (see text for details). At a distance of 2 kpc, 1 arcmin corresponds to 0.57 pc.} 
\label{fig:bow}
\end{figure*}

The SHS H$\alpha$+[N\,{\sc ii}] image presented in Fig.\,\ref{fig:bow} shows that besides the already known bow shock 
ahead of Vela\,X-1 (Kaper et al. 1997), there are also exist extended filamentary structures stretching in the downstream 
direction on both sides of the bow shock. The geometry of these filaments suggests that we see a wake left behind the 
supersonically moving system, which met on its way a wedge-like layer (or filament) of enhanced density. This suggestion 
(supported by the result of our 3D numerical modelling presented in Section\,\ref{sec:dis}) implies that the H$\alpha$ 
emission around Vela\,X-1 is a {\it density-bounded} \hii region and that the number density derived from 
equation\,(\ref{eqn:hii}) might be overestimated.   

Another way to estimate $n_{\rm ISM}$ is through the observed value of the stand-off distance of the bow shock (Baranov
et al. 1970):
\begin{equation}
R_0 =\left({\dot{M}v_\infty\over 4\pi\rho_{\rm ISM}v_{\rm rel} ^2}\right)^{1/2} \, ,
\label{eqn:st}
\end{equation}
where 
$\rho_{\rm ISM}=1.4m_{\rm H}n_{\rm ISM}$, $m_{\rm H}$ is the mass of a hydrogen atom, and $v_{\rm rel}$ is the stellar 
velocity relative to the local ISM (in the absence of regular flows in the ISM, $v_{\rm rel}=v_*$). Using the SHS image 
of the bow shock, we estimated $R_0$ to be 0.57 pc. With this $R_0$, $\dot{M}$ and $v_\infty$ from Section\,\ref{sec:vela}, 
and assuming that $v_{\rm rel}=v_*$, one finds
\begin{eqnarray}
n_{\rm ISM}\approx 1 \, {\rm cm}^{-3} \left({R_0\over 0.57 \, {\rm pc}}\right)^{-2}\left({v_*\over 54.3 \, \kms}\right)^{-2} 
\nonumber \\ \times \left({\dot{M}\over 6.3\times10^{-7} \, \myr}\right) \left({v_\infty \over 700 \, \kms}\right) \, .
\label{eqn:den}
\end{eqnarray}
If correct, this estimate\footnote{In Gvaramadze et al. (2011a), we obtained a factor of 10 larger value of $n_{\rm ISM}$ using 
the same arguments. The difference is mostly caused by factors of $\approx3.2$ and 1.6 larger values of $\dot{M}$ and 
$v_\infty$, respectively, adopted in that paper.} implies that the \hii region produced by Vela\,X-1 is actually a factor of 
several more extended than the {\it IRAS} source noted by Kaper et al. (1997). This in turn implies that Vela\,X-1 might be 
the main source of ionization of the elongated region of H$\alpha$ emission around it (see Fig.\,\ref{fig:22+shassa}).

\section{Spectroscopic observations of the bow shock
around Vela\,X-1}
\label{sec:obs}

To study the bow shock produced by Vela\,X-1, we obtained its spectrum with the Southern African Large Telescope (SALT; Buckley, 
Swart \& Meiring 2006; O'Donoghue et al. 2006) on 2014 March 16, using the Robert Stobie Spectrograph (RSS; Burgh et al. 2003; 
Kobulnicky et al. 2003) in the long-slit mode with a slit of 8 arcmin$\times$1.5 arcsec. The slit was placed to the west from 
Vela\,X-1 and oriented in the north-south direction (see the left-hand panel of Fig.\,\ref{fig:bow}). The PG900 grating was used to 
cover the spectral range of 4300--7400 \AA \, with a final reciprocal dispersion of 0.97 \AA \, pixel${-1}$ (FWHM$\approx$5.50 \AA). 
Three 900\,s spectra were taken with $2\times4$ binning. The RSS uses a mosaic of three 2048$\times$4096 CCDs and the final 
spatial scale for this observation was 0.5 arcsec  pixel$^{-1}$. The seeing during the observation was $\approx$0.8 arcsec. 
An Ar lamp arc spectrum was taken immediately after the science frame. A spectrophotometric standard star was observed during 
twilight time for relative flux calibration.

The primary reduction of the SALT data was carried out with the SALT science pipeline (Crawford et al. 2010). After that, the bias
and gain corrected and mosaicked long-slit data were reduced in the way described in Kniazev et al. (2008). Absolute flux calibration
is not feasible with SALT because the unfilled entrance pupil of the telescope moves during the observation. However, a relative
flux correction to recover the spectral shape was done using the observed spectrophotometric standard.

In the two-dimensional (2D) spectrum, the bow shock appears as emission lines of hydrogen Balmer series and forbidden lines of 
[O\,{\sc iii}], [N\,{\sc ii}], [S\,{\sc ii}], [Ar\,{\sc iii}] and [Ar\,{\sc iv}]. The emission is peaked at the position of two 
filaments, whose existence was already noted by Kaper et al. (1997). We label these filaments F1 and F2, as indicated in the 
left-hand panel of Fig.\,\ref{fig:bow}. The RSS slit crosses the filament F2 at the position close to that of the spectroscopic 
slit in the observations by Kaper et al. (1997; their region\,A). 

\begin{figure}
\includegraphics[width=6cm,angle=270,clip=]{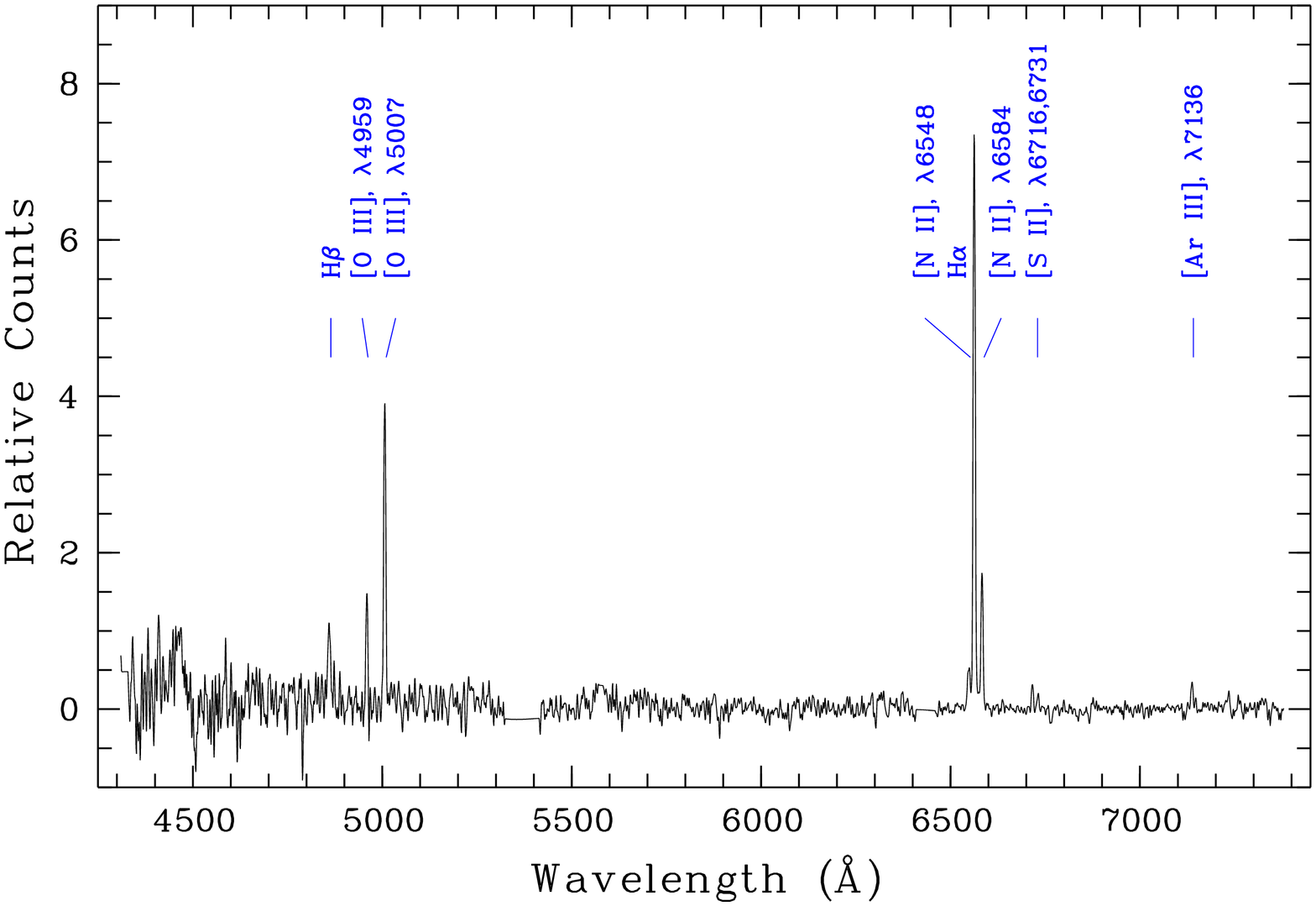}
\includegraphics[width=6cm,angle=270,clip=]{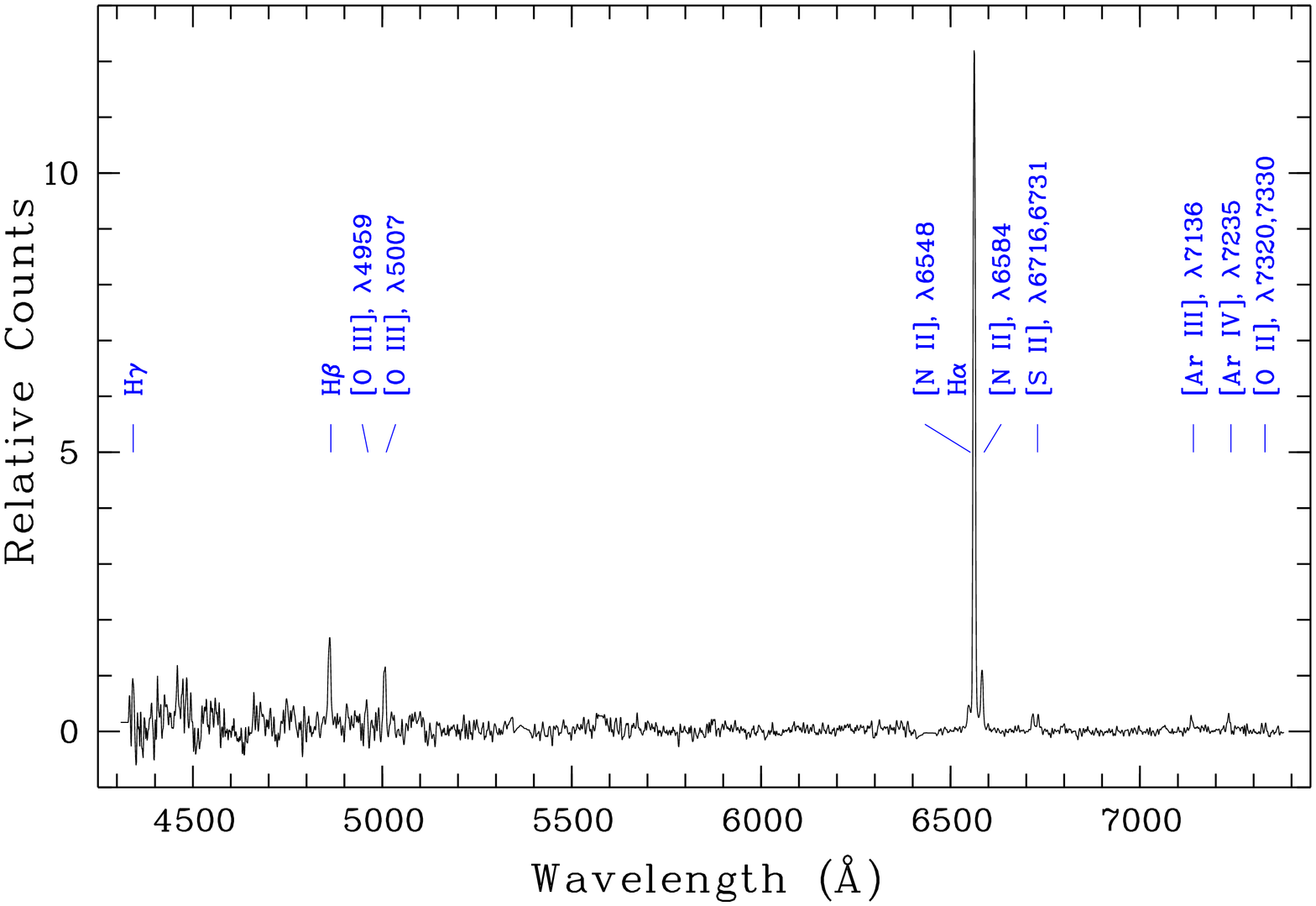}
\centering \caption{1D RSS spectra of the filaments F1 (upper panel) and F2 (bottom panel) in the bow shock around
Vela\,X-1. All detected emission lines are marked.} 
\label{fig:spec}
\end{figure}

One-dimensional (1D) spectra of F1 and F2 are shown in Fig.\,\ref{fig:spec}. The detected lines (measured with programs described 
in Kniazev et al. 2004) are listed in Table\,\ref{tab:int} along with their observed intensities (normalized to H$\beta$), 
$F(\lambda)/F({\rm H}\beta)$, the reddening-corrected line intensity ratios, $I(\lambda)/I({\rm H}\beta)$, the logarithmic 
extinction coefficients, $C({\rm H}\beta)$, and the colour excesses $E(B-V)$. Note that the H$\alpha$/H$\beta$ intensity ratio in 
the spectrum of F2 and the corresponding value of $E(B-V)$ are in good agreement with the values derived by Kaper et al. (1997)
for their region\,A. Compared to the 1.52-m telescope used by Kaper et al. (1997), the larger aperture of the SALT allowed us to 
detect the [S\,{\sc ii}] $\lambda\lambda$6717, 6731 lines in the spectra of both filaments and to estimate their electron number 
density, $n_{\rm e}$([S\,{\sc ii}]). The obtained densities are given in Table\,\ref{tab:int} as well. However, the weakness of the 
[S\,{\sc ii}] lines and correspondingly large errors in their measured intensities make the density estimates quite uncertain. On 
the other hand, our spectra do not show the [O\,{\sc i}] $\lambda$6300 line, which is present in one of the spectra obtained by 
Kaper et al. (1997). A possible explanation of this discrepancy is that this line might be the residual contamination due to 
improper subtraction of the night sky lines.

\begin{table}
\centering{
\caption{Line intensities of the filaments F1 and F2.}
\label{tab:int}
\begin{tabular}{lcc} \hline
 & F1 & \\
$\lambda_{0}$(\AA) Ion  & F($\lambda$)/F(H$\beta$)&I($\lambda$)/I(H$\beta$) \\ \hline
4861\ H$\beta$\         & 1.000$\pm$0.108 & 1.000$\pm$0.110 \\
4959\ [O\ {\sc iii}]\   & 0.854$\pm$0.107 & 0.821$\pm$0.104 \\
5007\ [O\ {\sc iii}]\   & 2.736$\pm$0.224 & 2.581$\pm$0.214 \\
6548\ [N\ {\sc ii}]\    & 0.465$\pm$0.050 & 0.264$\pm$0.030 \\
6563\ H$\alpha$\        & 5.105$\pm$0.398 & 2.888$\pm$0.247 \\
6584\ [N\ {\sc ii}]\    & 1.176$\pm$0.098 & 0.661$\pm$0.060 \\
6717\ [S\ {\sc ii}]\    & 0.181$\pm$0.032 & 0.098$\pm$0.018 \\
6731\ [S\ {\sc ii}]\    & 0.102$\pm$0.029 & 0.055$\pm$0.016 \\
7136\ [Ar\ {\sc iii}]\  & 0.201$\pm$0.033 & 0.099$\pm$0.017 \\
  & & \\
C(H$\beta$)                 & \MC {2}{l}{0.74$\pm$0.10} \\
$E(B$$-$$V$)                & \MC {2}{l}{0.50$\pm$0.07 mag}  \\
$n_{\rm e}$([S\,{\sc ii }]) & \MC {2}{l}{10$^{+850}_{-10}$ cm$^{-3}$}  \\
\hline
 & F2 & \\
 $\lambda_{0}$(\AA) Ion     & F($\lambda$)/F(H$\beta$)&I($\lambda$)/I(H$\beta$) \\ 
 \hline
4340\ H$\gamma$\           & 0.343$\pm$0.100 & 0.467$\pm$0.139 \\
4861\ H$\beta$\            & 1.000$\pm$0.181 & 1.000$\pm$0.184 \\
4959\ [O\ {\sc iii}]\      & 0.223$\pm$0.096 & 0.212$\pm$0.092 \\
5007\ [O\ {\sc iii}]\      & 0.628$\pm$0.134 & 0.580$\pm$0.126 \\
6548\ [N\ {\sc ii}]\       & 0.303$\pm$0.046 & 0.139$\pm$0.023 \\
6563\ H$\alpha$\           & 6.334$\pm$0.815 & 2.893$\pm$0.411 \\
6584\ [N\ {\sc ii}]\       & 0.669$\pm$0.090 & 0.303$\pm$0.045 \\
6717\ [S\ {\sc ii}]\       & 0.157$\pm$0.029 & 0.068$\pm$0.014 \\
6731\ [S\ {\sc ii}]\       & 0.138$\pm$0.029 & 0.059$\pm$0.013 \\
7136\ [Ar\ {\sc iii}]\     & 0.066$\pm$0.033 & 0.025$\pm$0.013 \\
7237\ [Ar\ {\sc iv}]\      & 0.110$\pm$0.045 & 0.040$\pm$0.017 \\
7320\ [O\ {\sc ii}]\       & 0.028$\pm$0.032 & 0.010$\pm$0.011 \\
7330\ [O\ {\sc ii}]\       & 0.037$\pm$0.041 & 0.013$\pm$0.015 \\
  & & \\
C(H$\beta$)                 & \MC {2}{l}{1.02$\pm$0.17} \\
$E(B$$-$$V$)                & \MC {2}{l}{0.69$\pm$0.12 mag}  \\
$n_{\rm e}$([S\,{\sc ii }]) & \MC {2}{l}{310$^{+960}_{-300}$ cm$^{-3}$}  \\
\hline
\end{tabular}
 }
\end{table}

For the sake of comparison with the results of numerical modelling, we estimated the H$\alpha$ surface brightness, 
$S_{{\rm H}\alpha}$, of the brightest filament (F2) in the bow shock at the position of the slit using the SHS image.
To do this, we used equations 1 and 2 in Frew et al. (2014) and the flux calibration factor of 9.5 counts pixel$^{-1}$ 
R$^{-1}$ from their table\,1, and adopted the observed [N\,{\sc ii}] to H$\alpha$ line intensity ratio of 0.15 from 
Table\,\ref{tab:int} (here [N\,{\sc ii}] is the sum of the $\lambda$6548 and $\lambda$6584 lines). We obtained the 
surface brightness corrected for the contribution from the contaminant [N\,{\sc ii}] lines of $S_{{\rm H}\alpha}\approx43$\,R, 
where 1 R$\equiv$1 Rayleigh =$5.66\times10^{-18} \, {\rm erg} \, {\rm s}^{-1} \, {\rm cm}^{-2} \, {\rm arcsec}^{-2}$ at 
H$\alpha$. 

\section{Numerical model}
\label{sec:mod}

The bow shock around Vela\,X-1 is modelled in the framework of a 3D non-stationary ideal-magnetohydrodynamic (MHD) 
model allowing us to take into account the interstellar and stellar magnetic fields. We apply a Godunov-type scheme that 
employs a Harten-Lax-van Leer Discontinuity (HLLD) MHD Riemann solver (see, e.g., Miyoshi \& Kusano 2005) and use a 3D moving 
grid with the possibility to fit the tangential discontinuity (or astropause) and the shocks. This is an important advantage 
of our model, because it allows us to significantly increase a quality of numerical calculations. The grid is non-uniform with 
increasing number of cells close to the discontinuities. In the region outside of the bow shock we also apply mesh refinement 
to set the dense layer. For further details on the numerical method, the reader is referred to Izmodenov \& Alexashov (2015).

In the framework of the model, we explored an interaction between the stellar wind of Vela\,X-1 and an inhomogeneous ISM 
(as specified below) in three limiting cases. In the first one (hereafter model\,1), we considered pure hydrodynamic flows. 
In the second simulation (model\,2), we considered the ISM with a regular, homogeneous magnetic field, while the stellar 
wind was assumed to be unmagnetized. In the third simulation (model\,3), we assumed that there is no magnetic field in the 
ISM and that the stellar wind possesses a helical magnetic field. The latter assumption is based on the following reasoning. 
The small orbital separation of the binary components in Vela\,X-1 and the high orbital velocity of the neutron star (of the 
same order magnitude as $v_\infty$) imply that there should be strong interaction between the wind of the blue supergiant and 
the magnetosphere of the neutron star, resulting in the origin of a helical magnetic field above and below of the orbital plane 
with the helix axis perpendicular to this plane. For the sake of simplicity, was assumed to this helical field is described 
by the Parker (1958) solution for the solar wind. The results of our simulations are presented and discussed in 
Section\,\ref{sec:dis}. Below, we present a general mathematical formulation of the problem.

We model the ISM and the stellar wind plasma as a one-component ideal fluid, governed by the following MHD equations:
\begin{eqnarray}
\label{eqn:mag1} &&
\frac{\partial \rho}{\partial t}  + \nabla\cdot(\rho{\boldsymbol v})=0,
\nonumber \\
&& \frac{\partial \rho
	{\boldsymbol v} }{\partial t}  + \nabla\cdot\left[\rho
{\boldsymbol v}{\boldsymbol v}+\left(p+\frac{B^2}{8\pi}\right) {\boldsymbol I}
-\frac{{\boldsymbol B}{\boldsymbol B}}{4\pi}\right] = 0,  \nonumber \\
&& \frac{\partial }{\partial t} \left(\epsilon+\frac{B^2}{8\pi}\right) + \nonumber \\ 
&&+\nabla\cdot\left[\left(\epsilon+p+\frac{B^2}{8\pi}\right){\boldsymbol v}
-\frac{({\boldsymbol v}\cdot{\boldsymbol B})}{4\pi}{\boldsymbol B}\right] = q , \\
&& \frac{\partial {\boldsymbol B}}{\partial t} + \nabla\cdot\left[{\boldsymbol v}{\boldsymbol B}-{\boldsymbol B}{\boldsymbol v}
\right]=0, \nonumber \\
&& \nabla\cdot{\boldsymbol B}= 0, \nonumber
\end{eqnarray}
where $\rho$, ${\boldsymbol v}$ and $p$ are, respectively, the plasma density, velocity and thermal pressure, ${\boldsymbol B}$ 
is the magnetic field, $\epsilon=\rho v^2/2 +p/(\gamma-1)$ is the kinetic plus internal energy per unit volume, $\gamma=5/3$ is
the adiabatic index, and ${\boldsymbol I}$ is the unity tensor.

The main difference of this model with that of Izmodenov and Alexashov (2015) is that here we take into account the radiative 
cooling by adding a source term to the energy equation of system (\ref{eqn:mag1}). The cooling term is $q = n_{\rm e}n_{\rm 
H}\Lambda(T)$, where $n_{\rm e}$ is the electron number density, $n_{\rm e}\approx n_{\rm H}$, $\Lambda(T)$ is the cooling 
function (adopted from Cowie et al. 1981), and $T$ is the temperature. 

\begin{figure*}
\includegraphics[width=18cm,clip=0]{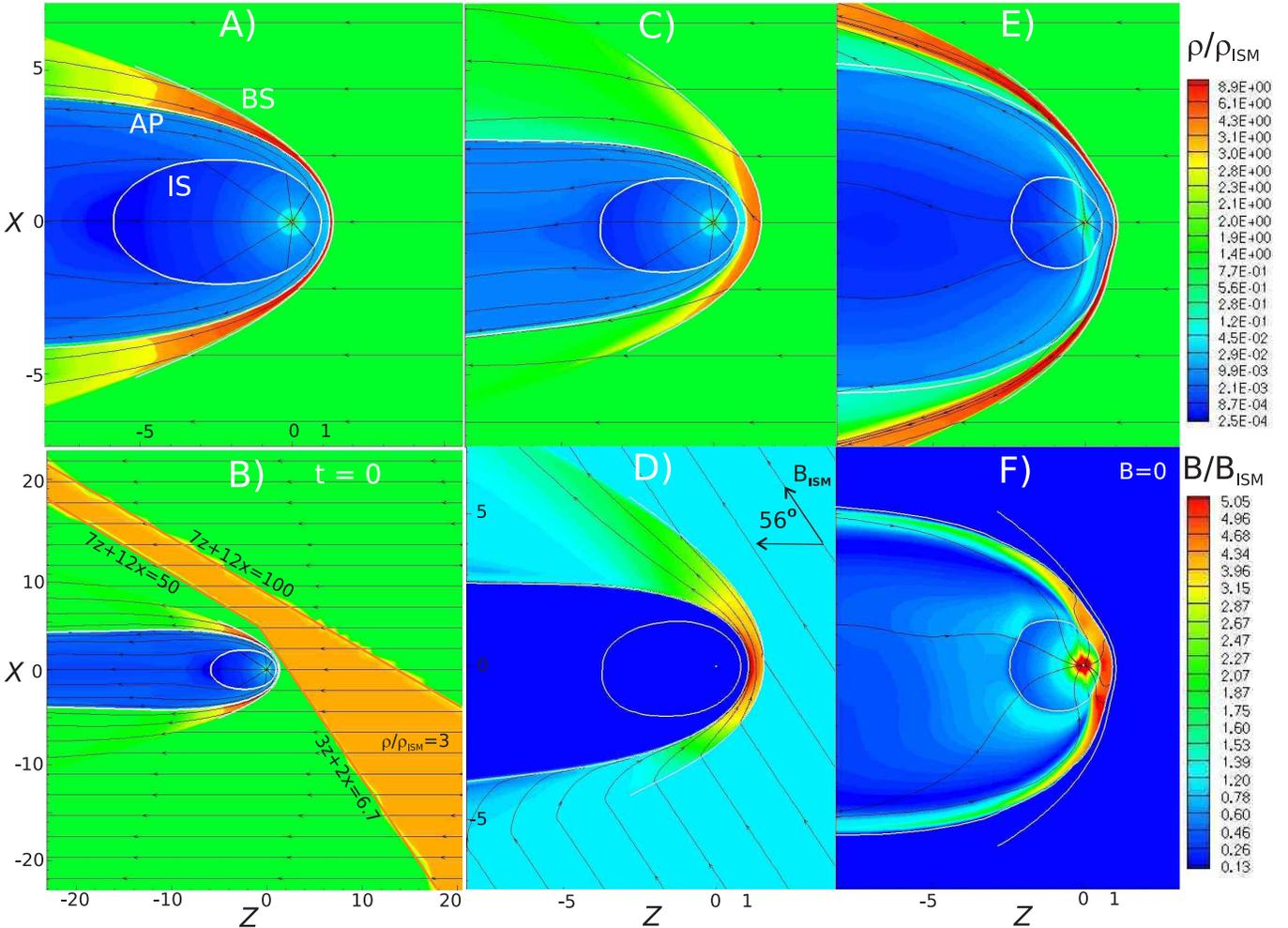}
\centering
\caption{2D distributions of the plasma density and streamlines (panels A, C an E) and the magnetic field with the field lines 
(panels D and F) in the steady-state models. Panel A corresponds to model\,1, panels C and D to model\,2, and panels E and F to 
model\,3. Panel\,B shows the initial condition in the ISM for the non-stationary models (see text for details). The ISM is 
flowing from right to left at $v_*$. The inner shock (IS), the astropause (AP) and the bow shock (BS) are plotted with white 
lines. The distance units on the X and Z axes are in $R_0=0.57$ pc.}
\label{fig:sta}
\end{figure*}

The inner boundary conditions were set at a distance $R_{\rm s}=0.025$ pc from the star. The stellar wind is characterized by the 
mass loss rate $\dot{M}$ and the terminal velocity $v_\infty$ (see Section\,\ref{sec:vela}). At $R_{\rm s}$ the stellar wind density 
is $\dot{M}/(4\pi v_\infty R_{\rm s}^2)$. Since the stellar wind is hypersonic, the thermal pressure in the 
wind was neglected. Note that the solution of the problem does not depend on the position of the inner boundary, 
provided that it is located inside the inner shock (see Fig.\,\ref{fig:sta}) and the stellar wind is hypersonic at this position.

In model\,3, we assumed that the magnetic field in the stellar wind is described by the Parker (1958) solution:
\begin{equation}
B_r=B_{\rm s}\left(\frac{R_{\rm s}}{r}\right)^2, B_{\theta_{\rm s}}=0, B_{\varphi_{\rm s}}=
2B_{\rm s}\left(\frac{R_{\rm s}}{r}\right) \, \sin\theta_{\rm s} \, ,
\end{equation}
where $B_r, B_{\theta_{\rm s}}$ and $B_{\varphi_{\rm s}}$ are the magnetic field components in the spherical coordinates, $\theta_{\rm s}$ 
is the angle counted from the axis perpendicular to the orbital plane of Vela\,X-1, and $\varphi_{\rm s}$ is the azimuthal angle. 
At large distances from the star $B_{\varphi_{\rm s}}$ dominates over other components. If Vela\,X-1 was produced by a spherically-symmetric 
supernova explosion, then the vector of its space velocity is confined in the orbital plane, and, correspondingly, the axis of the 
helical magnetic field is perpendicular to this velocity. To allow the possibility that the supernova explosion was
slightly asymmetric, we assumed that the angle between the magnetic field axis and the stellar velocity is equal to 80$\degr$
(we note that the results of the numerical simulations do not depend on the exact value of this angle, unless it significantly differs 
from 90$\degr$).

The outer boundary conditions were set in the undisturbed ISM with the number density $n_{\rm ISM}$, thermal pressure 
$p_{\rm ISM}=2n_{\rm ISM}kT_{\rm ISM}$, where $k$ is the Boltzmann constant and $T_{\rm ISM}$ is the temperature of the 
ISM (it was assumed to be equal to 8\,000\,K), and velocity ${\boldsymbol v}_{\rm ISM}=-{\boldsymbol v}_*$ (i.e. the modelling was 
performed in the reference frame of Vela\,X-1 and it was assumed that there is no large-scale motions in the ISM). In model\,2, we 
considered the effect of a uniform interstellar magnetic field, ${\boldsymbol B}_{\rm ISM}$, of strength of 5\,$\mu$G (which is typical 
of the Milky Way; e.g. Beck 2016) on the shape of the bow shock, and assumed that this field is inclined with respect to the ISM 
velocity vector by angle $\alpha$.

To simulate the density inhomogeneity in the local ISM (see Section\,\ref{sec:neb} and Fig.\,\ref{fig:22+shassa}), we
added to the problem a wedge-like layer of enhanced density $\rho=\beta \rho_{\rm ISM}$, where $\beta>1$ (see panel B of 
Fig.\,\ref{fig:sta}). To make the layer stable, we assumed that the thermal pressure in the layer is equal to that of the 
surrounding ISM (i.e. we did not consider heating of the layer by the radiation from Vela\,X-1). 

Formally, $n_{\rm ISM}$, $\beta$ and $B_{\rm s}$ can be considered as free parameters of the model.
On the other hand, the choice of these parameters determines the geometric structure of the bow shock (e.g. its curvature and 
stand-off distance) and its H$\alpha$ surface brightness, and therefore they could be constrained through the parametric study. 
In doing so, we found that the observations could be reproduced reasonably well if $n_{\rm ISM}=0.5$\,cm$^{-3}$, $\beta$=3 
and $B_{\rm s}=5\times10^{-4}$\,G. As expected, the obtained density is comparable to that given by equation (\ref{eqn:den}). 

To set the initial conditions for time-dependent models, we first obtained corresponding steady-state solutions (see 
Fig.~\ref{fig:sta} and Section\,\ref{sec:dis} for discussion of these solutions), where panels A, C and E plot distribution of 
the plasma density and streamlines, while panels D and F show the distribution of the magnetic field and field lines. We use the 
system of coordinates connected with the interstellar flow and ISM magnetic field vectors, ${\boldsymbol v}_{\rm ISM}$ and 
${\boldsymbol B}_{\rm ISM}$. The axis Z is directed toward the interstellar flow, i.e. opposite to ${\boldsymbol v}_{\rm ISM}$. 
The axis X is in the plane containing the ${\boldsymbol v}_{\rm ISM}$ and ${\boldsymbol B}_{\rm ISM}$ vectors and is perpendicular 
to the Z-axis. The axis Y completes a Cartesian right-hand system of coordinates. In this system, the wedge-like layer was assumed to
be infinite along the Y-axis. Also, for the sake of simplicity, we assumed that the axis of the stellar magnetic field is confined in 
the (XZ)-plane (we justify this assumption in Section\,\ref{sec:dis}).

At $t=0$, the bow shock generated by Vela\,X-1 was placed at a short distance from the wedge-like layer (see panel B of 
Fig.~\ref{fig:sta}). Then, the layer as a whole was set in motion towards Vela\,X-1 and after a while it started to interact with 
the bow shock. In our modelling, we performed calculations with different configurations of the layer, which was bounded by either 
two or three planes. This approach implies considerable freedom in the choice of the initial geometry of the layer (e.g. the 
separation between the planes and their orientations). Numerous trials showed that it is very difficult to choose a certain 
configuration which would allow ones to match the observations in their entirety. We therefore present the results obtained for 
one of the simplest configurations of the layer, which was assumed to be confined by three planes (as shown in panel B of 
Fig.\,\ref{fig:sta}). For the sake of definiteness, we assumed that ${\boldsymbol B}_{\rm ISM}$ is parallel to one of these planes 
(i.e. $\alpha=56\degr$; see panel D of Fig.\,\ref{fig:sta}). Despite of its simplicity, the adopted configuration of the layer 
allowed us to reproduce the main geometric features of the bow shock and the filamentary structures behind it, as well as their 
H$\alpha$ surface brightness. 

To compare the numerical models with the observations, we produced synthetic maps of H$\alpha$ emission. The surface brightness of 
the bow shock is an integral of the H$\alpha$ emissivity, $j_{{\rm H}\alpha}$, along the line of sight: $S_{{\rm H}\alpha}=\int 
j_{{\rm H}\alpha}(s)ds$, where 
\[
 j_{{\rm H}\alpha}(s)=2.85\times10^{-33} {\rm erg} \, {\rm s}^{-1} \, {\rm cm}^{-3} \, {\rm arcsec}^{-2} T_{\rm e} 
 ^{-0.9}(s)n_{\rm e} ^2(s) 
\]
(e.g. Mackey et al. 2013), $T_{\rm e}$ is the electron temperature and $s$ is the coordinate along the line of sight. The 
integration was performed only within the bow shock and the dense layer. The resulting synthetic maps in units of R are presented 
and discussed in the next section.

\section{Results and discussion}
\label{sec:dis}

\begin{figure*}
	\includegraphics[width=16cm,clip=0]{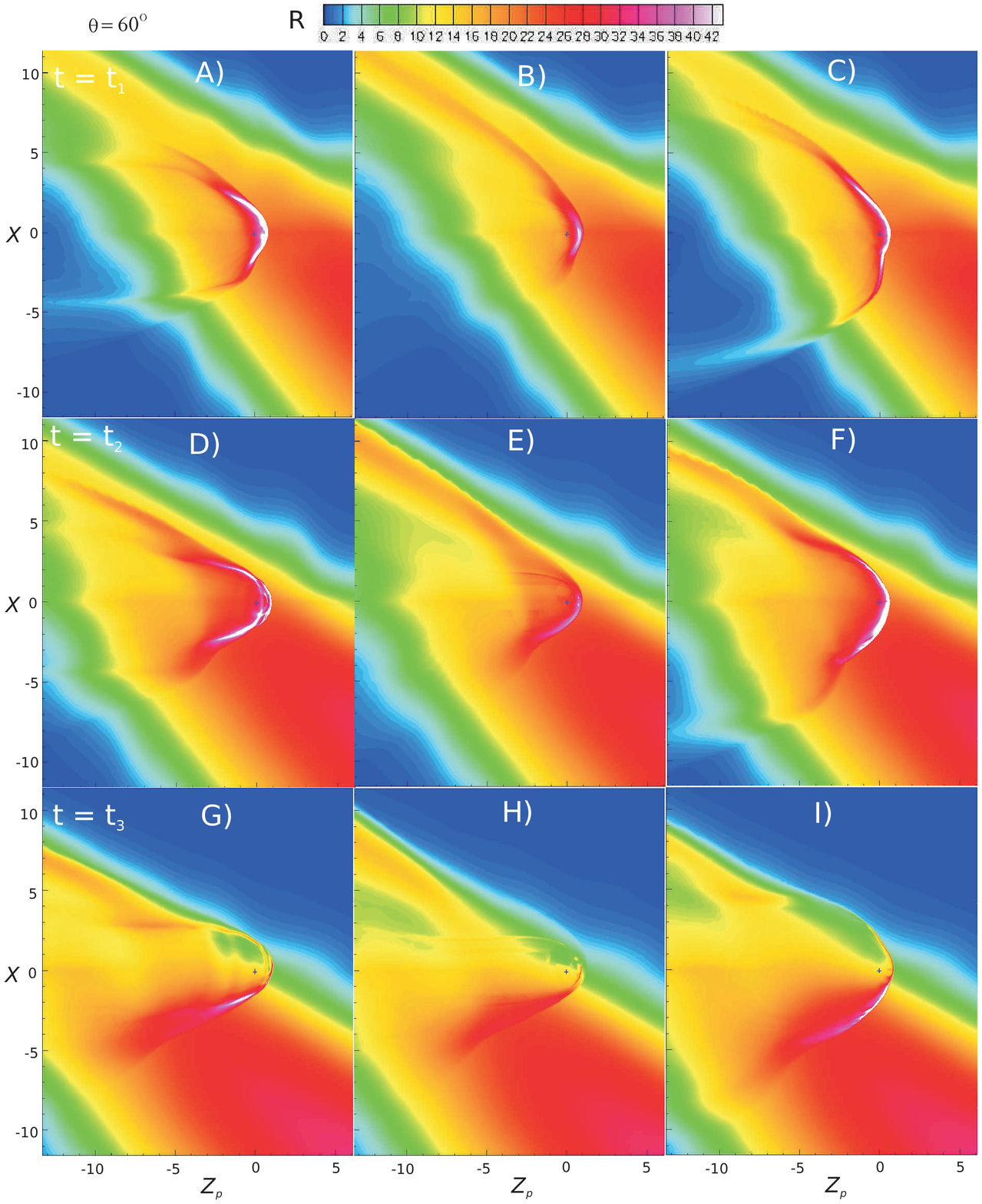}
	\centering \caption{
Projection of synthetic $H\alpha$ intensity maps (in R on the linear scale) with a line of sight at an angle of $\theta=60\degr$
to the (XZ)-plane of the non-stationary models\,1, 2 and 3 (left to right) at three times (top to bottom). The distance units 
on the X and Z$_{\rm p}$ axes are, respectively, in $R_0=0.57$ pc and $R_0\sin\theta$ (=0.49 pc).}
	\label{fig:ha}
\end{figure*}

Fig.\,\ref{fig:sta} plots distributions of the plasma density and streamlines (panels A, C an E) and the magnetic field with 
the field lines (panels D and F) for three steady-state models. The pure hydrodynamic model\,1 (panel\,A) is axisymmetric, while in 
models\,2 and 3 (respectively, panels C,D and E,F) the axial symmetry is broken by the magnetic field and the shape of the 
bow shocks becomes slightly asymmetric. Besides this, the main differences between the models are the opening angle of the bow shock
and the thickness of the layer of shocked ISM (bounded between the bow shock and the astropause). In model\,1, the shape of the bow 
shock (and astropause) is determined by the sonic Mach number, $M$, which depends only on the stellar velocity and the temperature 
of the local ISM (or the sound speed in this medium). For the adopted temperature of the ISM of 8\,000\,K, $M=4.3$. The panel\,A of 
Fig.\,\ref{fig:sta} shows that the opening angle of the bow shock in model\,1 is much smaller than the observed one. Indeed, the 
distance from the star to the model bow shock at an angle of 90$\degr$ from the direction of the stellar motion is $R(\pi/2)\approx1.8 
R_0$, while in the observed bow shock it is $R(\pi/2)\approx2.4 R_0$. In model\,2, the regular ISM magnetic field makes the bow shock 
more opened, $R(\pi/2)\approx2.1 R_0$, while the flanks of the astropause become more pressed to the symmetry axis as compared with 
model\,1. The inclusion of the stellar magnetic field in model\,3 makes both the astropause and the bow shock more opened than in 
model\,1, i.e. $R(\pi/2)\approx2.1 R_0$. The ISM magnetic field also significantly reduces compression and radiative cooling of the 
layer of the shocked ISM in model\,2, which makes this layer a factor of several thicker than in models\,1 and 3.

In the time-dependent models, the bow shock interacts with the approaching wedge-like layer of enhanced density and a complicated 
gasdynamic (or MHD) structure is formed in the downstream direction. For the parameters adopted in the simulations, the time for 
Vela\,X-1 to cross the simulation domain is 470\,000 yr. We splitted this time in 50 equal intervals and computed H$\alpha$ emission 
maps at the beginning of each interval. Fig.\,\ref{fig:ha} presents snapshots of the H$\alpha$ emission\footnote{Only a portion of the 
full simulation domain is shown.} at three moments of time, $t_1=93\,800$ yr, $t_2$=187\,600 yr and $t_3$=281\,400 yr, for a line of 
sight inclined by the angle $\theta=60\degr$ to the (XZ)-plane of the simulations (which corresponds to the inclination of the space 
velocity of Vela\,X-1 to our line of sight; see Section\,\ref{sec:vela}). Note that in this projection, the axis of the stellar helical 
magnetic field is inclined to our line of sight by an angle of 60$\degr$, which is close to the expected orientation of this axis given 
the observational fact that the orbital plane of Vela\,X-1 is inclined by an angle of $\ga70\degr$ with respect to the plane of sky 
(see Section\,\ref{sec:vela}).

Comparison of the H$\alpha$ maps at different time moments shows how the shape of the bow shock evolves in the course of interaction 
with the dense layer. One can see that at $t=t_2$ all models reproduce fairly well the general features in the wake behind Vela\,X-1. 
This supports our supposition that Vela\,X-1 is physically associated with the elongated region of H$\alpha$ emission around it and 
indicates that the filamentary structures stretched on both sides behind the bow shock are shaped by the stellar wind.

Fig.\,\ref{fig:ha} shows that the brightest arc of H$\alpha$ emission ahead of the star in all three models corresponds to the outer 
shock region, where the ISM is compressed and heated by the bow shock. Comparison of the models at $t=t_2$ shows that only in model\,3 
the opening angle of the arc agrees with the observations. To understand why in model\,2 this arc is less opened than in 
model\,3, while the opening angles of the bow shocks in both models are almost the same (see above), we recall that the ISM magnetic 
field reduces the degree of compression of the post-shock plasma, so that the plasma needs more time to cool down. As a result, 
the brightness of the resulting arc of H$\alpha$ emission in model\,2 is lower as compared with other two models, while the arc itself 
is well separated from the bow shock and, correspondingly, its opening angle is smaller than that of the bow shock. We performed 
calculations with different strengths and orientations of ${\boldsymbol B}_{\rm ISM}$, but failed to 
simultaneously match the observed opening angle and surface brightness of the bow shock in the framework of model\,2. Similarly, the high 
space velocity of Vela\,X-1 (and correspondingly the high ram pressure of the inflowing ISM) did not allow us to find conditions suitable 
to increase the opening angle of the pure hydrodynamic bow shock. These results motivated us to investigate the effect of the stellar 
magnetic field on the geometry of the bow shock.

Model\,3 not only better fits the opening angle of the bow shock and its H$\alpha$ brightness, but also allowed us to reproduce the 
apparent detachment of the eastern (left) wing of the bow shock from the wake (see the right-hand panel of Fig.\,\ref{fig:bow}) when 
the star reaches the outer edge of the layer at the time moment $t=t_3$ (cf. panel\,I 
of Fig.\,\ref{fig:ha}). Though at this moment the geometry of the wake has changed significantly, namely, the bend in its western 
(right) wing became less prominent and has shifted in the downstream direction, we expect that a better agreement with the observations 
would be achieved if the density in the layer grows towards the bottom right corner of panels of Fig.\,\ref{fig:ha}, which seems quite 
reasonable in view of the patchy appearance of the region of H$\alpha$ emission to the north-west of Vela\,X-1 (see 
Figs\,\ref{fig:22+shassa} and \ref{fig:bow}). 

\begin{figure}
	\includegraphics[width=8cm,clip=0]{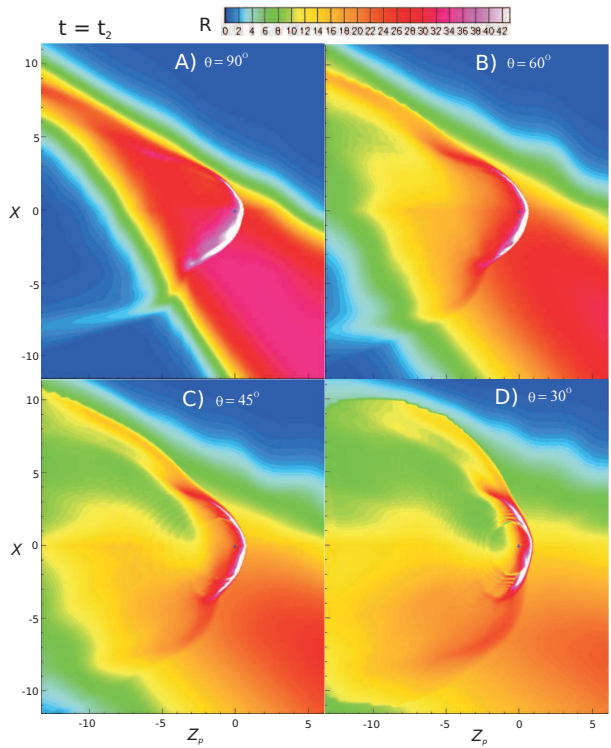}
	\centering \caption{Projection of synthetic $H\alpha$ intensity maps (in R on the linear scale) with a line of sight at 
	angles of $\theta=90\degr$, 60$\degr$, 45$\degr$ and 30$\degr$ to the (XZ)-plane of the non-stationary model\,2 and the time 
	moment $t_2$. The distance units on the X and Z$_{\rm p}$ axes are, respectively, in $R_0=0.57$ pc and $R_0\sin\theta$.}
	\label{fig:ham}
\end{figure}

In Fig.\,\ref{fig:ham}, we compare the H$\alpha$ maps obtained at $t=t_2$ from model\,3 for different inclinations of the (XZ)-plane
to a line of sight. One can see that the projections at the angles of $\theta=60\degr$ and 45$\degr$ provide an almost equally good 
agreement with the observations. Though the angle of 60$\degr$ is consistent with the orientation of the space velocity of Vela\,X-1 
(see Section\,\ref{sec:vela}) and therefore is preferable, the smaller angles cannot be exclude as well because the actual relative 
radial velocity of Vela\,X-1 with respect to the local ISM might be higher if the layer as a whole recedes from us in the radial 
direction. 

Although model\,3 reproduces reasonably well the opening and H$\alpha$ brightness of the bow shock around Vela\,X-1, as 
well as the main features of the wake behind this HMXB, there is room for further improvements of the model. In particular, our 
simulations do not consider the radiation transfer and its role in heating the local ISM and producing photoevaporation flows, which can 
affect the appearance of the bow shock. The inclusion of the radiative transfer is also necessary for proper modelling of the dust 
heating and ultimately for producing synthetic maps of infrared emission from the bow shock (cf. Mackey et al. 2016), and their comparison 
with images obtained by {\it Spitzer} (Iping, Sonneborn \& Kaper 2007; Gvaramadze et al. 2011a) and {\it Herschel}. 
This will be the subject of a future investigation.

Finally, for the sake of completeness, we briefly discuss the possible birth place of Vela\,X-1 and the effect of the bow shock on 
the structure of the supernova remnant, which, after some time, will be produced by the second supernova explosion in the system.

In Section\,\ref{sec:int}, we mentioned that van Rensbergen et al. (1996) and Kaper et al. (1997) suggested that Vela\,X-1 was born 
within the Vela\,OB1 association and then kicked-out from it because of the first supernova explosion in this massive binary system.
Their suggestion was based on available at that time low quality proper motion measurements and the orientation of the bow shock, 
both of which indicate that Vela\,X-1 is moving away from the outskirts of Vela\,OB1. The {\it Gaia} data (see Section\,\ref{sec:vela})
provide further support to this suggestion. 

\begin{figure}
\includegraphics[width=8.5cm,clip=0]{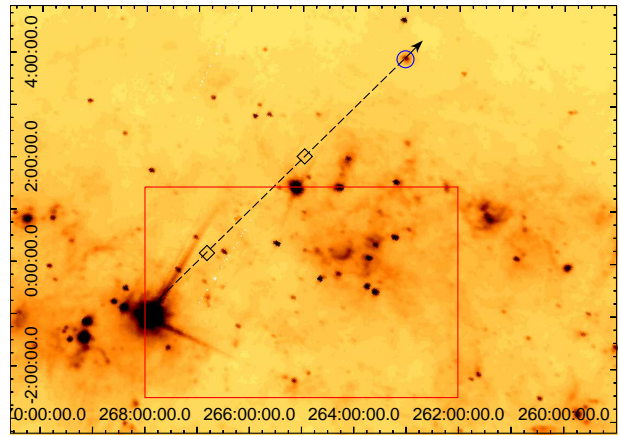}
\centering
\caption{{\it IRAS} 25\,$\mu$m image of the field containing Vela\,X-1 (marked by a circle) and the Vela\,OB1 association (whose 
approximate boundary is outlined by a red rectangle). The arrow shows the direction of motion of Vela\,X-1, while a dashed 
line indicates the trajectory of Vela\,X-1, as follows from the {\it Gaia} proper motion measurements. The diamonds on the 
trajectory of Vela\,X-1 show the positions of this HMXB 2 and 4 Myr ago. The bright source at $l\approx268\degr$ and 
$b\approx-1\degr$ is the young ($<1$ Myr), embedded star cluster RCW\,38 (Wolk, Bourke \& Vigil 2008). The coordinates (in 
units of degrees) are the Galactic longitude and latitude on the horizontal and vertical scales, respectively. At a distance 
of 2 kpc, 1$\degr$ corresponds to $\approx$34.4 pc. See text for details. }
\label{fig:ass}
\end{figure}

In Fig.\,\ref{fig:ass} we show the {\it Infrared Astronomical Satellite} ({\it IRAS}) 25\,$\mu$m image of the field containing 
Vela\,X-1 and the Vela\,OB1 association. The approximate boundary of Vela\,OB1 (Humphreys 1978) is outlined by a (red) rectangle. 
The dashed line shows the trajectory of Vela\,X-1 based on the {\it Gaia} observations. One can see that Vela\,X-1 is moving away 
from Vela\,OB1 and therefore this association indeed could be its birth place. We caution, however, that the relationship between 
Vela\,X-1 and Vela\,OB1 might, in fact, be spurious if the original binary system was ejected from its parent star cluster because 
of a few-body dynamical interaction during the early stage of cluster evolution and later on (when the system was already far from 
its birth place) attained an additional kick because of supernova explosion (Pflamm-Altenburg \& Kroupa 2010; Gvaramadze et al. 2012). 
In this case, the present space velocity of Vela\,X-1 is the vector sum of the ejection and kick velocities, and therefore it should not necessarily point away from the actual parent cluster.  

The high space velocity of Vela\,X-1 ($\sim50 \, \kms$) implies that by the moment of second supernova explosion (which 
will happened after several Myr from now) this HMXB will leave far behind the filamentary structures presented and discussed 
in this paper. Unless the system will meet another density inhomogeneity on its way, the shape of the resulting supernova remnant 
will be determined by the mass of the ISM accumulated within the bow shock (Brighenti \& D'Ercole 1994; Meyer et al. 2015), or 
ultimately on the number density of the local ISM and on whether or not the supernova exploded as a red supergiant or a Wolf-Rayet 
star (the current mass of the blue supergiant component of Vela\,X-1 of $24 \, \msun$ allows both possibilities). In both cases, 
the supernova remnant resulting from the interaction of the supernova blast wave with the bow shock material will appear as an 
incomplete circle with reduced brightness in the direction of the wake of the bow shock (e.g. Brighenti \& D'Ercole 1994; Meyer et al. 
2015). One cannot, however, exclude the possibility that at the end of its life Vela\,X-1 will fly into a superbubble on its way 
(cf. Huthoff \& Kaper 2002). In this case, the system will move at subsonic velocity though the hot, tenuous coronal gas and the 
bow shock will disappear, while the supernova remnant would be unobservable at all (e.g. Kafatos et al. 1980).

\section{Acknowledgements}

Numerical simulations and interpretation of the observational data were supported by the Russian Science Foundation grant No. 
14-12-01096. Spectroscopic observations with the SALT and data reduction, carried out by AYK, were supported by the National 
Research Foundation (NRF) of South Africa. This work is based in part on observations obtained with the Southern African 
Large Telescope (SALT), programme \mbox{2013-2-RSA\_OTH-003} and has made use of the NASA/IPAC Infrared Science 
Archive, which is operated by the Jet Propulsion Laboratory, California Institute of Technology, under contract with the National
Aeronautics and Space Administration, the Southern H-Alpha Sky Survey Atlas (SHASSA), which is supported by the National Science 
Foundation, and the SIMBAD data base and the VizieR catalogue access tool, both operated at CDS, Strasbourg, France.

\end{document}